\documentclass[twocolumn,showpacs,preprintnumbers,amsmath,amssymb,superscriptaddress]{revtex4}

\usepackage{multirow}
\usepackage{graphicx}
\usepackage{bm}

\newcommand\tabvspbot{\rule[-1.5ex]{0pt}{0pt}}

\begin{document}
\preprint{FLAVOUR(267104)-ERC-53, LTH 990,}
\preprint{SFB/CPP-13-82, TTP13-033}
\title{\boldmath $B_{s,d} \to \ell^+ \ell^-$~ in the Standard Model with 
                 Reduced Theoretical Uncertainty}
\author{Christoph~Bobeth}
\affiliation{Excellence Cluster Universe and TUM-IAS, Technische Universit\"at M\"unchen, D-85748 Garching, Germany}
\author{Martin~Gorbahn}
\affiliation{Department of Mathematical Sciences, University of Liverpool, Liverpool L69 3BX, United Kingdom}
\affiliation{Excellence Cluster Universe and TUM-IAS, Technische Universit\"at M\"unchen, D-85748 Garching, Germany}
\author{Thomas~Hermann}
\affiliation{Institut f\"ur Theoretische Teilchenphysik, Karlsruhe Institute of Technology (KIT), D-76128 Karlsruhe, Germany}
\author{Miko{\l}aj~Misiak}
\affiliation{Institute of Theoretical Physics, University of Warsaw, Ho\.za 69, PL-00-681 Warsaw, Poland}
\affiliation{Theory Division, CERN, CH-1211 Geneva 23, Switzerland}
\author{Emmanuel~Stamou}
\affiliation{Excellence Cluster Universe and TUM-IAS, Technische Universit\"at M\"unchen, D-85748 Garching, Germany}
\affiliation{Dept.\ of Particle Physics and Astrophysics, Weizmann Institute of Science, Rehovot 76100, Israel}
\author{Matthias~Steinhauser$\,$}
\affiliation{Institut f\"ur Theoretische Teilchenphysik, Karlsruhe Institute of Technology (KIT), D-76128 Karlsruhe, Germany}
\date{November 4, 2013}
\begin{abstract}
  We combine our new results for the ${\cal O}(\alpha_{em})$ and ${\cal
  O}(\alpha_s^2)$ corrections to $B_{s,d} \to \ell^+ \ell^-$, and
  present updated branching ratio predictions for these decays in the standard
  model. Inclusion of the new corrections removes major theoretical uncertainties 
  of perturbative origin that have just begun to dominate over the parametric ones.
  For the recently observed muonic decay of the $B_s$ meson, our calculation gives
  $\overline{\mathcal B}(B_s \to \mu^+ \mu^-) = (3.65 \pm 0.23) \times 10^{-9}$.
\end{abstract}

                             
\maketitle

Rare leptonic decays of the neutral $B$ mesons are highly suppressed in
the standard model (SM), and provide important constraints on models of new
physics. In the SM, these flavor changing neutral current decays are
generated first at one-loop level through W-box and Z-penguin diagrams. Their
branching ratios undergo an additional helicity suppression by
$m_\ell^2/M_{B_q}^2$, where $m_\ell$ and $M_{B_q}$ denote masses of the charged
lepton and the $B_q$ meson, respectively. This suppression can be lifted in
models with extra Higgs doublets, such as the minimal supersymmetric standard
model. Constraints on such models can be obtained even for the scalar masses
reaching a few TeV, far above the current direct search limits (see
e.g. Ref.~\cite{Arbey:2012ax}).  However, one of the key factors in
determining the constraints is the SM prediction accuracy. Improving this
accuracy is the main purpose of the present work.

The average time-integrated branching ratios
$\overline{\mathcal B}_{q\ell} \equiv \overline{\mathcal B}[B_q \to \ell^+ \ell^-]$
($q=s,d;~ \ell = e,\mu,\tau$) depend on details of $B_q\bar B_q$ mixing~\cite{DeBruyn:2012wk}.
A simple relation 
$\overline{\mathcal B}_{q\ell} = \Gamma[B_q \to \ell^+ \ell^-]/\Gamma^q_H$
holds in the SM to a very good approximation, with $\Gamma^q_H$
denoting the heavier mass-eigenstate total width.
%
%
For $\ell = \mu$, the current 
experimental world averages read~\cite{LHCb-CMS-combi:2013}
\begin{equation} \label{brexp}
{\overline{\mathcal B}_{s\mu}} = (2.9 \pm 0.7) \times 10^{-9},\hspace{5mm}
{\overline{\mathcal B}_{d\mu}} = \left(3.6^{+1.6}_{-1.4}\right) \times 10^{-10}.
\end{equation}
They have been obtained by combining the recent measurements of
CMS~\cite{Chatrchyan:2013bka} and LHCb~\cite{Aaij:2013aka}.  In the
${\overline{\mathcal B}_{s\mu}}$ case, reduction of uncertainties to a few
percent level is expected in the forthcoming decade. To match such an
accuracy, theoretical calculations must include the next-to-leading order
(NLO) corrections of electroweak (EW) origin, as well as QCD corrections up to
the next-to-next-to-leading order (NNLO). In the present paper, we combine our
new calculations of the NLO~EW~\cite{Bobeth:2013tba} and NNLO~QCD~\cite{Hermann:2013kca}
corrections to the relevant coupling constant (Wilson coefficient) $C_A$, and
present updated SM predictions for all the $\overline{\mathcal B}_{q\ell}$
branching ratios.

A convenient framework for describing the considered processes is an effective
theory derived from the SM by decoupling the top quark, the Higgs boson, and
the heavy electroweak bosons $W$ and $Z$ (see,
e.g., Ref.~\cite{Buras:1998raa} for a pedagogical introduction). The
effective weak interaction Lagrangian relevant for $B_q \to \ell^+ \ell^-$
reads
\begin{equation} \label{Lw}
{\mathcal L}_{\rm weak} = N\; C_A(\mu_b)\; 
(\bar b \gamma_\alpha \gamma_5 q)(\bar \ell \gamma^\alpha \gamma_5 \ell) + \ldots\,,
\end{equation}
where $C_A$ is the $\overline{\rm MS}$-renormalized Wilson coefficient at the
scale $\mu_b \sim m_b$. The ellipses stand for other, subleading weak
interaction terms (operators) which we discuss below.  The normalization
constant $N = V_{tb}^\star V_{tq}^{}\, G_F^2 M_W^2/ \pi^2$ is given in terms
of the Fermi constant $G_F$ (extracted from the muon decay), the $W$-boson
on-shell mass $M_W$, and the Cabibbo-Kobayashi-Maskawa (CKM) matrix elements
$V_{ij}$.

Once $C_A(\mu_b)$ is determined to sufficient accuracy, the branching ratio is
easily expressed in terms of the lepton mass $m_\ell$, the $B_q$-meson mass
$M_{B_q}$ and its decay constant $f_{B_q}$.  The latter is defined by the QCD
matrix element $\langle 0| \bar b \gamma^\alpha \gamma_5 q | B_q(p) \rangle = i p^\alpha f_{B_q}$.
One finds
\begin{equation} \label{br1}
\overline{\mathcal B}_{q\ell} \;=\; \frac{|N|^2 M_{B_q}^3 f_{B_q}^2}{8\pi\,\Gamma^q_H}\, 
\beta_{q\ell}\, r_{q\ell}^2\, |C_A(\mu_b)|^2 \,+\, {\mathcal O}(\alpha_{em}), 
\end{equation}
where $r_{q\ell} = 2 m_\ell/M_{B_q}$ and $\beta_{q\ell} =
\sqrt{1-r_{q\ell}^2}$. Equation~(\ref{br1}) holds at the leading order in
flavor-changing weak interactions and in $M_{B_q}^2/M_W^2$, which is 
accurate up to permille-level corrections. In particular, operators like
$(\bar b \gamma_5 q)(\bar \ell \ell)$ 
from 
the Higgs boson exchanges give rise to ${\mathcal
O}(M_{B_q}^2/M_W^2)$ effects only. Thus, one neglects such operators in the
SM. However, they often matter in beyond-SM theories.

As far as the ${\mathcal O}(\alpha_{em})$ term in Eq.~(\ref{br1}) is
concerned, it requires more explanation because we are going to neglect it
while including complete corrections of this order to $C_A(\mu_b)$. The first
observation to make is that some of the ${\mathcal O}(\alpha_{em})$
corrections to $C_A(\mu_b)$ get enhanced by $1/\sin^2\theta_W$, powers of
$m_t^2/M_W^2$ or logarithms $\ln^2 M_W^2/\mu_b^2$, as explained in
Ref.~\cite{Bobeth:2013tba}. None of these enhancements is possible for the ${\mathcal
O}(\alpha_{em})$ term in Eq.~(\ref{br1}) once $\mu_b \sim m_b$. This term is
$\mu_b$-dependent and contains contributions from operators like 
$(\bar b \gamma_\alpha \gamma_5 q)(\bar \ell \gamma^\alpha \ell)$
or
$(\bar b \gamma_\alpha P_L c)(\bar c \gamma^\alpha P_L s)$,
with photons connecting the quark and lepton lines. It depends on
non-perturbative QCD in a way that is not described by $f_{B_q}$ alone, and it
must compensate the $\mu_b$-dependence of $C_A(\mu_b)$.  Since we neglect this
term, scale dependence serves as one of the uncertainty estimates. When
$\mu_b$ is varied from $m_b/2$ to $2 m_b$, our results for $|C_A(\mu_b)|^2$
vary by about $0.3\%$, which corresponds to a typical size of ${\mathcal
O}(\alpha_{em})$ corrections that undergo no extra enhancement. On the other
hand, the NLO~EW corrections to $|C_A(\mu_b)|^2$ often reach a few percent
level~\cite{Bobeth:2013tba}.

\begin{figure}[t]
\includegraphics[width=8cm,angle=0]{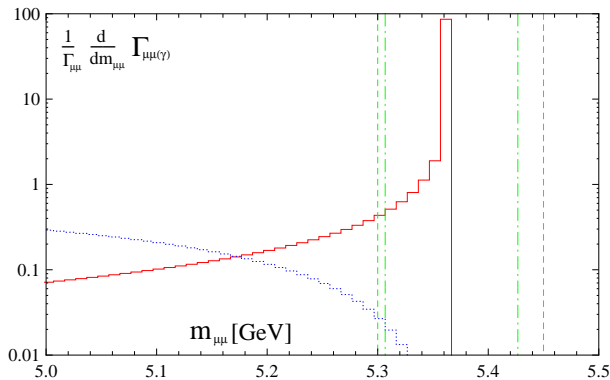}
\caption{\sf Contributions to the dimuon invariant-mass spectrum
in $B_s \to \mu^+ \mu^- (n\gamma)$ with $n=0,1,2,\ldots$ (see the text).
Both of them are displayed in bins of $0.01\,$GeV width.\label{fig:spectra}}
\end{figure}

The only other possible enhancement of QED corrections that one may worry
about is related to soft photon bremsstrahlung. For definiteness, let us
consider $B_s \to \mu^+ \mu^- (n\gamma)$ with $n=0,1,2,\ldots\,$.  The dimuon
invariant-mass spectrum in this process is obtained by summing the two
distributions shown in Fig.~\ref{fig:spectra}. The dotted (blue) curve
corresponds to real photon emission from the quarks (Eq.~(25) of
Ref.~\cite{Aditya:2012im}), while the tail of the solid (red) one is dominated
by soft photon radiation from the muons (Eqs.~(19)--(23) of
Ref.~\cite{Buras:2012ru}). The vertical dashed and dash-dotted (green) lines
indicate the CMS~\cite{Chatrchyan:2013bka} and LHCb~\cite{Aaij:2013aka} signal
windows, respectively.  In the displayed region below the windows
(i.e. between 5 and 5.3 GeV), each of the two contributions
integrates to around 5\% of the total rate.

The determination of $\overline{\mathcal B}_{s\mu}$ on the experimental side
includes a correction due to photon bremsstrahlung from the muons. For this
purpose, both CMS~\cite{Chatrchyan:2013bka} and LHCb~\cite{Aaij:2013aka} 
apply PHOTOS~\cite{Golonka:2005pn}. Such an approach is practically 
equivalent to extrapolating along the solid curve in Fig.~\ref{fig:spectra}
down to zero. In the resulting quantity, all the soft QED logarithms cancel
out, and we obtain $\overline{\mathcal B}_{s\mu}$ as in Eq.~(\ref{br1}), up to
${\mathcal O}(\alpha_{em})$ terms that undergo no extra
enhancement~\cite{Buras:2012ru}.

The direct emission, i.e. real photon emission from the quarks is infrared
safe by itself because the decaying meson is electrically neutral. It is
effectively treated as background on both the experimental and theoretical
sides. On the experimental side, it is neglected in the signal window (being
very small there, indeed), and not included in the extrapolation. On the theory
side, it is just excluded from $\overline{\mathcal B}_{s\mu}$ by definition.
This contribution survives in the limit $m_\mu \to 0$, which explains its
considerable size below the signal window in Fig.~\ref{fig:spectra}.
\begin{table}[t]
\begin{center}
\begin{tabular}{llcl}
  Parameter
& Value
& Unit
& Ref.
\\
\hline
  $G_F$                   
& $1.166379 \times 10^{-5}$
& GeV$^{-2}$
& \cite{Beringer:1900zz} 
\\
  $\alpha_s^{(5)}(M_Z)$
& $0.1184\, (7)$
& --
& \cite{Beringer:1900zz}	
\\
  $\alpha_{em}^{(5)}(M_Z)$
& $1/127.944\, (14)$
& --
& \cite{Beringer:1900zz}	
\\
  $\Delta \alpha_{em, {\rm hadr}}^{(5)}(M_Z)$
& $0.02772\, (10)$
& --
& \cite{Beringer:1900zz}	
\\
  $M_Z$ 
& $91.1876\, (21)$ 
& GeV
& \cite{Beringer:1900zz}	
\\
  $M_t$
& $173.1\, (9)$
& GeV
& \cite{Beringer:1900zz}	
\\
\tabvspbot
  $M_H$
& $125.9 \,(4)$
& GeV
& \cite{Beringer:1900zz}	
\\
\hline
  $M_{B_s}$
& $5366.77\, (24)$
& MeV
& \cite{Beringer:1900zz} 
\\
  $M_{B_d}$
& $5279.58\, (17)$
& MeV
& \cite{Beringer:1900zz}
\\
  $f_{B_s}$
& $227.7\,(4.5)$
& MeV
& \cite{Aoki:2013ldr} 
\\
  $f_{B_d}$
& $190.5\,(4.2)$
& MeV
& \cite{Aoki:2013ldr} 
\\
  $1/\Gamma^s_H$
& $1.615\,(21)$
& ps
& \cite{Amhis:2012bh}
\\
\tabvspbot
  $2/(\Gamma^d_H+\Gamma^d_L)$
& $1.519\,(7)$
& ps
& \cite{Amhis:2012bh}
\\
\hline
  $|V_{cb}|$
& $0.0424\,(9)$
& --
& \cite{Gambino:2013rza}
\\
  $|V_{tb}^\star V_{ts}^{}/V_{cb}^{}|$
& $0.980\,(1)$
& --
& \cite{Charles:2004jd,Ciuchini:2000de}
\\
\tabvspbot
  $|V_{tb}^\star V_{td}^{}|$
& $0.0088\,(3)$
& --
& \cite{Charles:2004jd,Ciuchini:2000de}                  
\\
\hline
\end{tabular}
\caption{\sf Numerical inputs. \label{tab:numinput} }
\end{center}
\end{table}

In this context, one may wonder whether the helicity suppression factor
$r^2_{q\ell}$ in Eq.~(\ref{br1}) can be relaxed at higher orders in QED. For
the two-body decay it is not possible in the SM because a generic
non-local interaction of $B_q$ with massless leptons contains vector or
axial-vector lepton currents contracted with the lepton momenta, which means
that it vanishes on shell. On the other hand, contributions with (real or virtual) 
%
%
photons coupled to the quarks may survive in the $m_\ell \to 0$
limit, but they are phase-space suppressed in the signal window (cf. the
dotted line in Fig.~\ref{fig:spectra}). In the $\overline{\mathcal B}_{s\mu}$
case, the phase-space suppression is at least as effective as the helicity
suppression, given the applied window sizes in both experiments.

We are now ready to numerically evaluate the branching ratios in
Eq.~(\ref{br1}). Our inputs are collected in Table~\ref{tab:numinput}. The
$\overline{\rm MS}$-renormalized coupling constants $\alpha_s^{(5)}(M_Z)$ and
$\alpha_{em}^{(5)}(M_Z)$ are defined in the SM with decoupled top
quark. Hadronic contributions to the evolution of $\alpha_{em}$ are given by
$\Delta\alpha_{em, {\rm hadr}}^{(5)}$.  This quantity is used to evaluate the
$W$-boson pole mass according to the fit formula in Eqs.~(6) and (9) of
Ref.~\cite{Awramik:2003rn}, which gives $M_W = 80.358\,(8)\,$GeV,
consistently with the direct measurement $M_W =
80.385\,(15)\,$GeV~\cite{Beringer:1900zz}. All the masses in
Table~\ref{tab:numinput} are interpreted as the on-shell ones.  In the 
top-quark case, this is equivalent to assuming that the so-called color
reconnection effects are included in the uncertainty. Converting $M_t$ to the
$\overline{\rm MS}$-renormalized mass with respect to QCD (but still on shell
with respect to EW interactions), we get $m_t \equiv m_t(m_t) = 163.5\,$GeV.

The decay constants $f_{B_q}$ are adopted from the most recent update of the
$N_f = (2 + 1)$ FLAG compilation~\cite{Aoki:2013ldr} which
averages the $N_f=2+1$ results of Refs.~\cite{Bazavov:2011aa, McNeile:2011ng, Na:2012kp}.
More recent calculations with $N_f=2+1+1$~\cite{Dowdall:2013tga}  and
$N_f=2$~\cite{Carrasco:2013zta} are consistent with these averages.
As far as the lifetimes are concerned, using the explicit result for $\tau^s_H
\equiv 1/\Gamma^s_H$ from Ref.~\cite{Amhis:2012bh} allows to avoid considering
correlations between the decay width difference and the average lifetime. In
the case of $B_d$, we can safely set $1/\Gamma^d_H \simeq
2/(\Gamma^d_H+\Gamma^d_L) \equiv \tau^d_{\rm av}$ given the tiny SM
expectation for $(\Gamma^d_L-\Gamma^d_H)/(\Gamma^d_L+\Gamma^d_H)\, 
\equiv \Delta \Gamma^d/(2\Gamma^d_{\rm av}) =
0.0021\,(4)$~\cite{Lenz:2011ti}.

The CKM matrix element $|V_{cb}|$ is treated in a special manner, as it is now
responsible for the largest parametric uncertainty in $\overline{\mathcal
B}_{s\mu}$. One should be aware of a long-lasting tension between its
determinations from the inclusive and exclusive semileptonic 
decays~\cite{Aoki:2013ldr}. Here, we adopt the recent inclusive fit
from Ref.~\cite{Gambino:2013rza}. It is the first one where both the
semileptonic data and the precise quark mass determinations from
flavor-conserving processes have been taken into account. Once $|V_{cb}|$ is
fixed, we evaluate $|V_{tb}^\star V_{ts}^{}|$ using the accurately known ratio
$|V_{tb}^\star V_{ts}^{}/V_{cb}^{}|$.

Apart from the parameters listed in Table~\ref{tab:numinput}, our results
depend on two renormalization scales $\mu_0 \sim M_t$ and $\mu_b \sim m_b$
used in the calculation of the Wilson coefficient $C_A$. This dependence is
very weak thanks to our new calculations of the NLO~EW and NNLO~QCD
corrections. Since this issue is discussed at length in the parallel
articles~\cite{Bobeth:2013tba,Hermann:2013kca}, we just fix here these scales to
$\mu_0 = 160\,$GeV and $\mu_b = 5\,$GeV.  Our results for the Wilson
coefficient $C_A$ are then functions of the first seven parameters in
Table~\ref{tab:numinput}. Allowing only the top-quark mass and the strong
coupling constant to deviate from their central values, we find the
following fits for $C_A$
\begin{eqnarray} 
%
C_A(\mu_b)  &=& 0.4802\; R_t^{ 1.52}\; R_\alpha^{-0.09}\; 
              - 0.0112\; R_t^{  0.89}\; R_\alpha^{-0.09}\nonumber\\[1mm]
            &=& 0.4690\; R_t^{  1.53}\; R_\alpha^{-0.09}\,,\label{cAfit1} \\[2mm]
%
C_A(\mu_b)  &=& 0.4802\; \widetilde{R}_t^{ 1.50}\; R_\alpha^{ 0.015}\; 
              - 0.0112\; \widetilde{R}_t^{0.86}\; R_\alpha^{-0.031}\nonumber\\[1mm]
            &=& 0.4690\; \widetilde{R}_t^{ 1.51}\; R_\alpha^{0.016}\,,\label{cAfit2} 
\end{eqnarray}
where $R_\alpha = \alpha_s(M_Z)/0.1184$,~ $R_t = M_t/(173.1\,{\rm GeV})$ and
$\widetilde{R}_t = m_t/(163.5\,{\rm GeV})$. The fits are accurate to
better than $0.1\%$ in $C_A$ for $\alpha_s(M_Z) \in [0.11,\, 0.13]$, $M_t
\in [170,\, 175]\,$GeV, and $m_t \in [160,\, 165]\,$GeV.

In the first lines of Eqs.~(\ref{cAfit1}) and~(\ref{cAfit2}), $C_A$ is given
as as a sum of two terms. The first one corresponds to the leading order EW
but NNLO QCD matching calculation~\cite{Hermann:2013kca}.  The second one accounts
for the NLO EW matching corrections~\cite{Bobeth:2013tba} at the scale $\mu_0$,
as well as for the logarithmically enhanced QED corrections that originate
from the renormalization group evolution between $\mu_0$ and
$\mu_b$~\cite{Bobeth:2003at,Huber:2005ig}.

Inserting Eq.~(\ref{cAfit1}) into Eq.~(\ref{br1}), 
we obtain for $\overline{\mathcal{B}}_{s\mu}$
\begin{equation}
\overline{\mathcal{B}}_{s\mu} \times 10^9 = (3.65 \pm 0.06)\, R_{t\alpha}\, R_s
= 3.65 \pm 0.23, \label{bsmu}
\end{equation}
where $R_{t\alpha} = R_t^{  3.06}\; R_\alpha^{-0.18} 
                       = \widetilde{R}_t^{  3.02}\; R_\alpha^{0.032}$ and 
\begin{displaymath}
R_s = \left( \frac{f_{B_s}[{\rm MeV}]}{227.7} \right)^{\! 2}\!
      \left( \frac{|V_{cb}|}{0.0424} \right)^{\! 2}\!
      \left( \frac{|V_{tb}^\star V_{ts}/V_{cb}|}{0.980} \right)^{\! 2}
             \frac{\tau_H^s\,[{\rm ps}]}{1.615}\,.
\end{displaymath}
Correlations between $f_{B_s}$ and $\alpha_s$ have been ignored above.
Uncertainties due to parameters that do not occur in the quantities
$R_\alpha$, $R_t$ and $R_s$ have been absorbed into the residual error
in the middle term of Eq.~(\ref{bsmu}). This residual error is actually
dominated by a non-parametric uncertainty, which we set to $1.5\%$ of the
branching ratio. Such an estimate of the non-parametric uncertainty is
supposed to include:
\begin{itemize}
\item[(i)] Effects of the neglected ${\mathcal O}(\alpha_{em})$ term in
        Eq.~(\ref{br1}).  They account for the fact that
        $|C_A(\mu_b)|^2$ changes by around $0.3\%$ when $\mu_b$ is varied
        between $m_b/2$ and $2\,m_b$. Such a dependence on $\mu_b$ must cancel
        order-by-order in perturbation theory.
\item[(ii)] Higher-order ${\mathcal O}(\alpha_s^3, \alpha_{em}^2, \alpha_s
        \alpha_{em})$  matching corrections to $C_A$ at the electroweak scale
        $\mu_0$.  Such corrections must remove the residual $\mu_0$-dependence
        of $C_A(\mu_b)$.  When $\mu_0$ is varied between $m_t/2$ and $2\,m_t$,
        the variation of $|C_A(\mu_b)|^2$ due to EW and QCD interactions
        amounts to around $0.2\%$ in each
        case~\cite{Bobeth:2013tba,Hermann:2013kca}. Effects of similar size in the
        branching ratio are observed in Ref.~\cite{Bobeth:2013tba} when comparing 
        several EW renormalization schemes.
\item[(iii)] Higher-order ${\mathcal O}(M_{B_q}^2/M_W^2)$ power corrections.
\item[(iv)] Uncertainties due to evaluation of $m_t$ from the experimentally
            determined $M_t$ using a three-loop relation. Note that
            half of the three-loop correction shifts $m_t$ by about
            $200\,$MeV, which affects $\overline{\mathcal{B}}_{s\mu}$ by
            around $0.3\%$. Non-perturbative uncertainties at this point
            (renormalons, color reconnection) are expected to be of the same
            order of magnitude.
\item[(v)] Tiny ${\mathcal O}(\Delta\Gamma^q/\Gamma^q)$ corrections due to
deviations from the relation $\overline{\mathcal B}_{q\ell} = \Gamma[B_q \to
\ell^+ \ell^-]/\Gamma^q_H$, i.e. due to decays of the lighter mass
eigenstate in the $B_q\bar B_q$ system. At the leading order in $\alpha_{em}$
and $M_{B_q}^2/M_W^2$, such corrections are non-vanishing only because of
CP-violation in the absorptive part of the $B_q\bar B_q$ mixing matrix. Apart
from being suppressed by $\Delta\Gamma^q/\Gamma^q$, they vanish in the
limit $m_c \to m_u$, and receive additional CKM suppression in the $B_s$
case. Beyond the leading order in $\alpha_{em}$ or $M_{B_q}^2/M_W^2$, the
lighter eigenstate can decay to leptons also in the CP-conserving limit of the
SM.
\end{itemize}

All the other $\overline{\mathcal{B}}_{q\ell}$ branching ratios are calculated 
along the same lines. We find
\begin{eqnarray}
\overline{\mathcal{B}}_{se} \times 10^{14} &=& (8.54 \pm 0.13)\, R_{t\alpha}\, R_s
= 8.54 \pm 0.55,\nonumber\\
\overline{\mathcal{B}}_{s\tau} \times 10^7\; &=& (7.73 \pm 0.12)\, R_{t\alpha}\, R_s
= 7.73 \pm 0.49,\nonumber\\
\overline{\mathcal{B}}_{de} \times 10^{15} &=& (2.48 \pm 0.04)\, R_{t\alpha}\, R_d
= 2.48 \pm 0.21,\nonumber\\
\overline{\mathcal{B}}_{d\mu} \times 10^{10} &=& (1.06 \pm 0.02)\, R_{t\alpha}\, R_d
= 1.06 \pm 0.09,\nonumber\\
\overline{\mathcal{B}}_{d\tau} \times 10^8\; &=& (2.22 \pm 0.04)\, R_{t\alpha}\, R_d
= 2.22 \pm 0.19,~~~~~ \label{bql.other}
\end{eqnarray}
with
\begin{displaymath}
R_d = \left( \frac{f_{B_d}[{\rm MeV}]}{190.5} \right)^2
      \left( \frac{|V_{tb}^\star V_{td}|}{0.0088} \right)^2
             \frac{\tau_d^{\rm av}\,[{\rm ps}]}{1.519}\,.
\end{displaymath}
A summary of the error budgets for $\overline{\mathcal{B}}_{s\ell}$ and
$\overline{\mathcal{B}}_{d\ell}$ is presented in
Table~\ref{tab:BR:results}. It is clear that the main parametric uncertainties
come from $f_{B_q}$ and the CKM angles.

To get rid of such uncertainties, one may take advantage~\cite{Buras:2003td}
of their cancellation in ratios like
\begin{equation} \label{kappa}
\kappa_{q\ell} \,\equiv\,
\frac{\overline{\mathcal{B}}_{q\ell}\, \Gamma_H^q\, \Delta M_{B_q}^{-1}}{(G_F M_W m_\ell)^2 \beta_{q\ell}}
\,\stackrel{\scriptscriptstyle\rm SM}{\simeq}\, 
\frac{ 3 \,|C_A(\mu_b)|^2}{\pi^3\, C_{LL}(\mu_b)\, B_{B_q}(\mu_b)}\, ,
\end{equation}
where $\Delta M_{B_q}$ is the mass difference in the $B_q\bar B_q$ system, and
$C_{LL}$ enters through the $\Delta B=2$ term in ${\mathcal L}_{\rm weak}$, namely
$\, -\frac14 N\, V_{tb}^\star V_{tq}^{}\, C_{LL} (\bar b \gamma_\alpha P_L q)(\bar b\gamma^\alpha P_L q)$.
The bag parameters $B_{B_q}$ are defined by the QCD matrix elements 
$\langle \bar B_q| (\bar b \gamma_\alpha P_L q)(\bar b\gamma^\alpha P_L q) |B_q \rangle 
= \frac23 f^2_{B_q} B_{B_q} M_{B_q}^2$.

Following FLAG~\cite{Aoki:2013ldr}, we take $\hat B_{B_s} = 1.33(6)$ and
$\hat B_{B_d} = 1.27(10)$~\cite{Gamiz:2009ku}. For the Wilson coefficient
$C_{LL}$, including the NLO QCD~\cite{Buras:1990fn} and NLO
EW~\cite{Gambino:1998rt} corrections, we find $\hat C_{LL} \equiv
C_{LL}(\mu_b)\, B_{B_q}(\mu_b)/\hat B_{B_q} = 1.27\; R_t^{1.51}$ for
$\alpha_s^{(5)}(M_Z) = 0.1184$ and $\mu_b = 5\,$GeV.  The r.h.s.\ of
Eq.~(\ref{kappa}) gives then $\kappa_{s\ell} = 0.0126(7)$ and
$\kappa_{d\ell} = 0.0132(12)$.  It follows that the overall
theory uncertainties in $\kappa_{q\ell}$ and $\overline{\mathcal{B}}_{q\ell}$
are quite similar at present.  The l.h.s.\ of Eq.~(\ref{kappa}) together with
Eq.~(\ref{brexp}) give $\kappa_{s\mu}^{\rm exp} = 0.0104(25)$ and
$\kappa_{d\mu}^{\rm exp} = 0.047(20)$, which is consistent with the SM
predictions.
\begin{table}[t]
\begin{center}
\renewcommand{\arraystretch}{1.4}
\begin{tabular}{c|ccc|cc|c|c|c}
& $f_{B_q}$ 
& CKM 
& $\tau_H^q$ 
& $M_t$ 
& $\alpha_s$ 
& other      
& non-        
& $\sum$
\\[-2mm]
& & & & & & param. & param. & 
\\
\hline
  $\overline{\mathcal{B}}_{s\ell}$
& $4.0$\%
& $4.3\%$ 
& $1.3$\%
& $1.6$\%
& $0.1$\%
& $< 0.1$\%
& $1.5$\%
& $6.4$\%
\\
  $\overline{\mathcal{B}}_{d\ell}$
& $4.5$\%
& $6.9$\%
& $0.5$\%
& $1.6$\%
& $0.1$\%
& $< 0.1$\%
& $1.5$\%
& $8.5$\%
\end{tabular}
\renewcommand{\arraystretch}{1.0}
\caption{ \label{tab:BR:results} 
\sf Relative uncertainties from various sources in $\overline{\mathcal{B}}_{s\ell}$ 
and $\overline{\mathcal{B}}_{d\ell}$. In the last column they are added in quadrature.}
\end{center}
\end{table}

To conclude, we have presented updated SM predictions for all the
$\overline{\mathcal{B}}_{q\ell}$ branching ratios. Thanks to our new results
on the NLO~EW~\cite{Bobeth:2013tba} and NNLO~QCD~\cite{Hermann:2013kca}
matching corrections, a significant reduction of the non-parametric
uncertainties has been achieved. Such uncertainties are now estimated at the
level of around 1.5\% of the branching ratios, compared to around 8\%
prior to our calculations. As far as the parametric ones are concerned,
their reduction will depend on progress in the lattice determinations of
$f_{B_q}$ and $B_{B_q}$ in the cases of $\overline{\mathcal{B}}_{q\ell}$ and
$\kappa_{q\ell}$, respectively.  For $\overline{\mathcal{B}}_{q\ell}$, the CKM
uncertainties are now equally important, with $|V_{cb}|$ being one of the main
limiting factors in the precise determination of
$\overline{\mathcal{B}}_{s\ell}$.

The increased theory accuracy is essential in interpreting the
experimental findings in terms of the SM or new physics. This will be
particularly important after the LHCb upgrade (see
e.g. Ref.~\cite{Alessio:2013ima}), when the experimental accuracy in
${\overline{\mathcal B}_{s\mu}}$ is expected to reach the same level as the
current theoretical one.  Even if no deviation from the SM is found, the role
of $B_q \to \ell^+ \ell^-$ in constraining new physics will become
significantly stronger.

\begin{acknowledgments}
We would like to thank Andrzej Buras, Gino Isidori, Paul Rakow, Jochen
Schieck and Zbigniew W{\c{a}}s for helpful discussions. This work was supported by the DFG through
the SFB/TR~9 ``Computational Particle Physics'' and the Graduiertenkolleg
``Elementarteilchenphysik bei h\"ochster Energie und h\"ochster
Pr\"azision''. C.B. has been supported in part by the ERC project
``FLAVOUR'' (267104). M.G. acknowledges partial support by the UK Science
\& Technology Facilities Council (STFC) under grant No. ST/G00062X/1. The work
of M.M. has been partially supported by the National Science Centre (Poland)
research project, Decision No. DEC-2011/01/B/ST2/00438.  %
\end{acknowledgments}

\end{document}